\documentclass[a4paper,12pt,reqno]{amsart}
\usepackage{amssymb}
\usepackage{amsfonts}
\usepackage{txfonts}
\usepackage{amsmath}
\usepackage{verbatim}
\usepackage{graphicx}
\usepackage{hyperref}
\usepackage{amsthm}
\usepackage{amsxtra}
\usepackage{amscd}
\usepackage{latexsym}

\newtheorem{theorem}{Theorem}[section]
\newtheorem{proposition}[theorem]{Proposition}
\newtheorem{lemma}[theorem]{Lemma}

\newtheorem{remark}[theorem]{Remark}

\renewenvironment{proof}[1][Proof]{\textbf{#1.} }{\ \rule{0.5em}{0.5em}}
\renewcommand{\theequation}{\thesection.\arabic{equation}}

\begin{document}
\title[Inhomogeneous parabolic equations]{Inhomogeneous parabolic equations
on unbounded metric measure spaces}
\author[Falconer]{Kenneth J. Falconer}
\address{Mathematical Institute, University of St Andrews, North Haugh, St
Andrews, Fife KY16 9SS, UK.}
\email{kjf@st-andrews.ac.uk}
\author[Hu]{Jiaxin Hu}
\address{Department of Mathematical Sciences, Tsinghua University, Beijing
100084, China.}
\email{hujiaxin@mail.tsinghua.edu.cn}
\author[Sun]{Yuhua Sun}
\address{Department of Mathematical Sciences, Tsinghua University, Beijing
100084, China.}
\email{sunyh08@mails.tsinghua.edu.cn}
\thanks{ JH was supported by NSFC (Grant No. 11071138).\\
KJF thanks the Department of Mathematical Sciences, Tsinghua University, for
their hospitality}

\begin{abstract}
We study inhomogeneous semilinear parabolic equations with source term $f$
independent of time $u_{t}=\Delta u+u^{p}+f(x)$ on a metric measure space,
subject to the conditions that $f(x)\geq 0$ and $u(0,x)=\varphi (x)\geq 0$.
By establishing Harnack-type inequalities in time $t$ and some powerful
estimates, we give sufficient conditions for non-existence, local existence,
and global existence of weak solutions. This paper generalizes previous
results on Euclidean spaces to general metric measure spaces.
\end{abstract}

\maketitle
\tableofcontents

\renewcommand{\theequation}
{\arabic{section}.\arabic{equation}} \numberwithin{equation}{section}


\section{Introduction}

\label{SecIntr} \setcounter{equation}{0} In recent years, the study of PDEs
on self-similar fractals has attracted increasing interest, see for example
\cite{Dalry, Fal99, FalHu01, Kig01, STR06}. In this paper we investigate a
class of nonlinear diffusions with source terms on general metric measure
spaces. Diffusion is of fundamental importance in many areas of physics,
chemistry, and biology. Applications of diffusion include: (1) Sintering,
i.e. making solid materials from powder (powder metallurgy, production of
ceramics); (2) Catalyst design in the chemical industry; (3) Steel can be
diffused (e.g. with carbon or nitrogen) to modify its properties; (4) Doping
during production of semiconductors; (5) The well-known Black-Scholes Model
in Financial Mathematics that is closely related to option pricing can be
transformed to a parabolic equation.

Let $\left( M,d,\mu \right) $ be a \emph{metric measure} space, that is, $%
\left( M,d\right) $ is a locally compact separable metric space and $\mu $
is a Radon measure on $M$ with full support. We consider the following
nonlinear diffusion equation with a source term $f$ on $(M,d,\mu )$:
\begin{equation}
u_{t}=\Delta u+u^{p}+f(x),\;t>0\text{ and }x\in M,  \label{eq}
\end{equation}%
with initial value
\begin{equation}
u(0,x)=\varphi (x),  \label{eq:init}
\end{equation}%
where $p>1$ and $f,\varphi :M\rightarrow \mathbb{R}$ are non-negative
measurable functions. With an appropriate interpretation of weak solutions
of (\ref{eq}) on $(M,d,\mu )$, we shall investigate the non-existence (or
blow-up) of solutions, the local and global existence of weak solutions to (%
\ref{eq})-(\ref{eq:init}), as well as the regularity of these solutions.
Although we were partially motivated by a series of earlier papers \cite%
{Bandle00,FalHu01,Fujita66, GHL03, Wei80, Wei81,zhang98}, there are new
ideas in this paper. In particular, we have used the theory of heat kernels
on metric measure spaces.

Recall the definition of the heat kernel which will be central to our
approach. A function $k(\cdot ,\cdot ,\cdot ):\mathbb{R}_{+}\times M\times
M\rightarrow \mathbb{R}$ is called a \emph{heat kernel} if the following
conditions $(k1)-(k4)$ are fulfilled: for $\mu $-almost all {$(x,y)\in
M\times M$ and for all } $t,s>0,$

\begin{enumerate}
\item[$(k1)$] \textit{Markov property}{:\ $k(t,x,y)>0$, and $%
\int_{M}k(t,x,y)d\mu (y)\leq 1$;}

\item[$(k2)$] \textit{symmetry}{:\ $k(t,x,y)=k(t,y,x)$;}

\item[$(k3)$] \textit{semigroup property}{:\ $k(s+t,x,z)=%
\int_{M}k(s,x,y)k(t,y,z)d\mu (y)$;}

\item[$(k4)$] \textit{normalization}{: for all $f\in L^{2}(M,\mu )$}%
\begin{equation*}
{\lim\limits_{t\rightarrow 0^{+}}\int_{M}k(t,x,y)f(y)d\mu (y)=f(x)}\text{ in
the }{L^{2}(M,\mu )}\text{{-norm.}}
\end{equation*}
\end{enumerate}

We assume that the heat kernel $k(t,x,y)$ considered in this paper is
jointly continuous in $x,y,$ and hence the above formulae in $(k1)-(k4)$
hold for \emph{every} {$(x,y)\in M\times M$. }

Two typical examples of heat kernels in $\mathbb{R}^{n}$ are the
Gauss-Weierstrass and the Cauchy-Poisson kernels:
\begin{eqnarray*}
k(t,x,y) &=&\frac{1}{(4\pi t)^{n/2}}\exp {\left( -\frac{|x-y|^{2}}{4t}%
\right) }, \\
k(t,x,y) &=&\frac{C_{n}}{t^{n}}\left( 1+\frac{|x-y|^{2}}{t^{2}}\right)
^{-(n+1)/2}\text{ \ }\left( C_{n}=\frac{\Gamma \big({\textstyle\frac{1}{2}}%
(n+1)\big)}{\pi ^{(n+1)/2}}\right) .
\end{eqnarray*}%
Jointly continuous sub-Gaussian heat kernels exist on many basic fractals,
for example, on the Sierp\'{\i}nski gasket, see Barlow and Perkins \cite%
{BP88}, and on Sierp\'{\i}nski carpets, see Barlow and Bass \cite{BB99,
Bar98}. For other fractals see \cite{HK99, Kig01}. For non-sub-Gaussian heat
kernels, see \cite{BBCK09, ChenK08}.

A heat kernel $k$ is called \emph{conservative} if it satisfies

\begin{enumerate}
\item[$(k5)$] \textit{conservative property:}{\ $\int_{M}k(t,x,y)d\mu (y)=1$%
, for all $t>0$ and all $x\in M$.}
\end{enumerate}

We will also assume that the heat kernel satisfies the following estimates

\begin{enumerate}
\item[$(k6)$] \textit{two-sided bounds}{: there exist constants }$\alpha
,\beta >0$ such that for all $t>0$ and all $x,y\in M,${\ \
\begin{equation}
\frac{1}{t^{\alpha /\beta }}\Phi _{1}\left( \frac{d(x,y)}{t^{1/\beta }}%
\right) \leq k(t,x,y)\leq \frac{1}{t^{\alpha /\beta }}\Phi _{2}\left( \frac{%
d(x,y)}{t^{1/\beta }}\right)  \label{hk-est}
\end{equation}%
w}here $\Phi _{1}$ and $\Phi _{2}$ are strictly positive and non-increasing
functions on $[0,\infty )$.
\end{enumerate}

It turns out that the parameter $\alpha $ in (\ref{hk-est}) is the \emph{%
fractal dimension}, and $\beta $ is the \emph{walk dimenison} of $M$, see
\cite{GHL03}.

Two-sided estimates (\ref{hk-est}) hold on various fractals where
\begin{equation*}
\Phi _{i}(s)=C_{i}\exp (-c_{i}s^{\beta /(\beta -1)})\text{ \ }(\text{for all
}s\geq 0)
\end{equation*}%
for constants $C_{i},c_{i}>0$ $(i=1,2)$ and $\beta >2$ is the walk dimension$%
.$

To prove the regularity of solutions, we need to assume that the heat kernel
$k$ is H\"{o}lder continuous in the space variables:

\begin{enumerate}
\item[$(k7)$] \textit{H\"{o}lder continuity}{:\ there exist constants $L>0$,
$\nu \geq 1$ and $0<\sigma \leq 1$ such that}%
\begin{equation*}
{\ }\left\vert {k(t,x_{1},y)-k(t,x_{2},y)}\right\vert {\leq Lt^{-\nu
}d(x_{1},x_{2})^{\sigma }}
\end{equation*}%
f{or all $t>0$ and all $x_{1},x_{2},y\in M$.}
\end{enumerate}

Given a heat kernel $k$, the operator $\Delta $ in (\ref{eq}) is interpreted
as the \emph{infinitesimal generator} of the \emph{heat semigroup} $\left\{
K_{t}\right\} _{t\geq 0}$ in $L^{2}:=L^{2}(M,\mu )$. Thus we let
\begin{equation}
K_{t}g(x)=\int_{M}k(t,x,y)g(y)d\mu (y)\ (t>0,g\in L^{2}),  \label{semi:1}
\end{equation}%
and define $\Delta $ by%
\begin{equation}
\Delta g=\lim_{t\downarrow 0}\frac{K_{t}g-g}{t}\text{ (in }L^{2}\text{-norm).%
}  \label{generaror}
\end{equation}%
Observe that $\{K_{t}\}_{t>0}$ is a strongly continuous and contractive
semigroup in $L^{2}$, that is, for all $s,t\geq 0$ and all $g\in L^{2},$
\begin{eqnarray}
K_{s+t} &=&K_{s}K_{t},  \label{semi:2} \\
\lim_{t\rightarrow 0^{+}}||K_{t}g-g||_{2} &=&0,  \notag \\
||K_{t}\phi ||_{q} &\leq &||\phi ||_{q}\text{ \ (for all }1\leq q\leq \infty
\text{).}  \notag
\end{eqnarray}%
The domain of the operator $\Delta $ is dense in $L^{2}.$

A function $u(t,x)$ is termed a \emph{weak solution} to (\ref{eq})-(\ref%
{eq:init}) if it satisfies the following integral equation
\begin{equation}
u(t,x)=K_{t}\varphi (x)+\int_{0}^{t}K_{\tau }f(x)d\tau
+\int_{0}^{t}K_{t-\tau }u^{p}(\tau ,x)d\tau ,  \label{weak solution}
\end{equation}%
where $K_{t}$ is the heat semigroup defined in (\ref{semi:1}).

The structure of this paper is as follows. In Section \ref{non}, we show the
non-existence of weak solution to (\ref{eq})-(\ref{eq:init}). In Section \ref%
{exis}, we obtain sufficient conditions for the local and global existence
of solutions for a range of parameters $p$, source terms $f$ and intial
values $\varphi $. The critical exponents $p$ depend only on the fractal
dimension $\alpha $ and the walk dimension $\beta $. Finally, in Section \ref%
{regular}, we investigate the H\"{o}lder continuity of weak solutions.

\textbf{Notation}\emph{.} The letters $C,C_{i}(i=1,2,\ldots )$ denote
positive constants whose values are unimportant and may differ at different
occurrences.

\section{Non-existence of solutions}

\label{non} In this section we give sufficient conditions for the
non-existence of essentially bounded solutions. The exponents $p=1+\beta
/\alpha $ (where $\alpha ,\beta >0)$, and $p=\alpha /(\alpha -\beta )$,
where $\alpha >\beta >0$ occur in the heat kernel bounds (\ref{hk-est}),
play a crucial r\^{o}le in our analysis, see Theorem \ref{Non-existence}.
First, we establish Lemma \ref{Harnack}, where condition $(k6)$ is our only
assumption on the heat kernel $k$ (we do not need the conservative property
of $k$ at this stage.)

The following properties the functions $\Phi _{1}$ and $\Phi _{2}$ in
condition $(k6)$ may or may not hold: there exist positive constants $%
a_{i},b_{i}$ and $c_{i}$ such that, for all $s,t\geq 0,$
\begin{eqnarray}
\Phi _{1}(s) &\geq &a_{1}\Phi _{2}(a_{2}s),  \label{general1} \\
\Phi _{2}(s+t) &\geq &b_{1}\Phi _{2}(b_{2}s)\Phi _{2}(b_{3}t),
\label{general4} \\
\Phi _{1}^{p}(s) &\geq &c_{1}\Phi _{2}(c_{2}s).  \label{general2}
\end{eqnarray}%
Note that if (\ref{general1}) holds, then $0<a_{1}\leq 1$ by letting $s=0$
and using the fact that $\Phi _{2}(0)\geq \Phi _{1}(0).$ Without loss of
generality, we may assume that $a_{2}>1$ in (\ref{general1}), since if (\ref%
{general1}) holds for some $a_{2}\leq 1,$ it also holds for any constant $%
a_{2}>1$ by the monotonicity of $\Phi _{2}.$

The \emph{Gauss-type} functions $\Phi _{1}$ and $\Phi _{2}$
\begin{eqnarray}
\Phi _{1}(s) &=&C_{1}\exp (-C_{2}s^{\gamma }),  \notag \\
\Phi _{2}(s) &=&C_{3}\exp (-C_{4}s^{\gamma }),\;s\geq 0,  \label{G-func}
\end{eqnarray}%
for constants $\gamma >0$ and $C_{i}>0 \,(1\leq i\leq 4)$ satisfy properties
(\ref{general1})-(\ref{general2}). The \emph{Cauchy-type} functions
\begin{eqnarray}
\Phi _{1}(s) &=&C_{1}\left( 1+s\right) ^{-\gamma },  \notag \\
\Phi _{2}(s) &=&C_{2}\left( 1+s\right) ^{-\gamma },\;s\geq 0  \label{C-func}
\end{eqnarray}%
for constants $\gamma >0$ and $C_{i}>0\,(i=1,2)$, satisfy properties (\ref%
{general1}) and (\ref{general4}), but not (\ref{general2}) if $p>1.$

Condition $(k6)$ and inequality (\ref{general1}) lead to the following key
lemma.

\begin{lemma}
\label{Harnack} Assume that the heat kernel $k$ satisfies condition $(k6)$
and (\ref{general1}). Then, for all non-negative measurable functions $g$ on
$M$ and for all $t>0,x\in M,$
\begin{eqnarray}
K_{t}g(x) &\geq &A_{1}K_{Bt}g(x),  \label{harnack1} \\
\int_{0}^{t}K_{\tau }g(x)d\tau &\geq &A_{2}tK_{B^{2}t}g(x),  \label{harnack2}
\end{eqnarray}%
where $A_{1}=a_{1}a_{2}^{-\alpha }<1,A_{2}=a_{1}a_{2}^{-2\alpha
}(1-a_{2}^{-\beta })<1$ and $B=a_{2}^{-\beta }<1.$ Consequently, for all
non-negative measurable functions $\varphi ,$
\begin{equation}
K_{t}\varphi (x)+\int_{0}^{t}K_{\tau }g(x)d\tau \geq A\left[
K_{B_{1}t}\varphi (x)+tK_{B_{^{1}}t}g(x)\right] ,  \label{eqn-3}
\end{equation}%
where $A=\min \left\{ A_{1}^{2},A_{2}\right\} <1$ and $B_{1}=B^{2}=a_{2}^{-2%
\beta }.$
\end{lemma}

\begin{proof}
It follows from condition $(k6)$ and (\ref{general1}) that
\begin{eqnarray}
K_{t}g(x) &=&\int_{M}k(t,x,y)g(y)d\mu (y)  \notag \\
&\geq &\int_{M}\frac{1}{t^{\alpha /\beta }}\Phi _{1}\left( \frac{d(x,y)}{%
t^{1/\beta }}\right) g(y)d\mu (y)  \notag \\
&\geq &a_{1}\int_{M}\frac{1}{t^{\alpha /\beta }}\Phi _{2}\left( a_{2}\frac{%
d(x,y)}{t^{1/\beta }}\right) g(y)d\mu (y).  \label{eqn-1}
\end{eqnarray}%
which gives that, using $(k6)$ again,
\begin{eqnarray*}
K_{t}g(x) &\geq &a_{1}a_{2}^{-\alpha }\int_{M}k\left( a_{2}^{-\beta
}t,x,y\right) g(y)d\mu (y) \\
&=&a_{1}a_{2}^{-\alpha }K_{a_{2}^{-\beta }t}g(x)=A_{1}K_{Bt}g(x),
\end{eqnarray*}%
proving (\ref{harnack1}).

To show (\ref{harnack2}), we see from (\ref{eqn-1}) that for all $\tau \in
\lbrack a_{2}^{-\beta }t,t],$ using the monotonicity of $\Phi _{2}$ and
condition $(k6),$
\begin{eqnarray*}
K_{\tau }g(x) &\geq &a_{1}\int_{M}\frac{1}{\tau ^{\alpha /\beta }}\Phi
_{2}\left( a_{2}\frac{d(x,y)}{\tau ^{1/\beta }}\right) g(y)d\mu (y) \\
&\geq &a_{1}\int_{M}\frac{1}{t^{\alpha /\beta }}\Phi _{2}\left( a_{2}\frac{%
d(x,y)}{(a_{2}^{-\beta }t)^{1/\beta }}\right) g(y)d\mu (y) \\
&\geq &a_{1}a_{2}^{-2\alpha }\int_{M}k\left( a_{2}^{-2\beta }t,x,y\right)
g(y)d\mu (y) \\
&=&a_{1}a_{2}^{-2\alpha }K_{B^{2}t}g(x).
\end{eqnarray*}%
Therefore,%
\begin{eqnarray*}
\int_{0}^{t}K_{\tau }g(x)d\tau &\geq &\int_{a_{2}^{-\beta }t}^{t}K_{\tau
}g(x)d\tau \\
&\geq &\int_{a_{2}^{-\beta }t}^{t}a_{1}a_{2}^{-2\alpha }K_{B^{2}t}g(x)d\tau
\\
&=&a_{1}a_{2}^{-2\alpha }\left( 1-a_{2}^{-\beta }\right) tK_{B^{2}t}g(x),
\end{eqnarray*}%
proving (\ref{harnack2}).

Finally, replacing $t$ by $Bt,$ we see from (\ref{harnack1}) that $%
K_{Bt}\varphi (x)\geq A_{1}K_{B^{2}t}\varphi (x)=A_{1}K_{B_{1}t}\varphi (x),$
and thus
\begin{equation}
K_{t}\varphi (x)\geq A_{1}K_{Bt}\varphi (x)\geq AK_{B_{1}t}\varphi (x).
\label{eqn-2}
\end{equation}%
Adding (\ref{harnack2}) and (\ref{eqn-2}), we obtain (\ref{eqn-3}).
\end{proof}

Lemma \ref{Harnack} gives the following estimate (\ref{nonexistence-bound})
that plays an important r\^{o}le in proving the non-existence of global
bounded solutions.

\begin{theorem}
\label{Non-existence} Assume that the heat kernel $k$ satisfies conditions $%
(k6)$ and (\ref{general1}). Let $u(t,x)$ be a non-negative essentially
bounded solution of (\ref{weak solution}) in $(0,T)\times M$. Then, for all $%
(t,x)\in (0,T)\times M,$
\begin{equation}
t^{1/(p-1)}K_{B_{1}t}\varphi (x)+t^{p/(p-1)}K_{B_{1}t}f(x)\leq C_{1},
\label{nonexistence-bound}
\end{equation}%
where $B_{1}=a_{2}^{-2\beta }$ as before, and $C_{1}$ depends only on $p$
(and in particular is independent of $T,\varphi $ and $f$).
\end{theorem}

\begin{proof}
Observe that by condition $(k1)$ and using a weighted H\"{o}lder inequality$%
, $ for all $t>0,x\in M$ and for all non-negative functions $g,$
\begin{eqnarray*}
K_{t}\left( g^{p}\right) (x) &=&\int_{M}k(t,x,y)g^{p}(y)d\mu (y) \\
&\geq &\left[ \int_{M}k(t,x,y)g(y)d\mu (y)\right] ^{p}=\left[ K_{t}g(x)%
\right] ^{p}.
\end{eqnarray*}%
It follows from (\ref{weak solution}) and (\ref{eqn-2}) that
\begin{eqnarray}
u(t,x) &\geq &\int_{0}^{t}K_{t-\tau }u^{p}(\tau ,x)d\tau  \notag \\
&\geq &A\int_{0}^{t}K_{B_{1}(t-\tau )}u^{p}(\tau ,x)d\tau  \notag \\
&\geq &A\int_{0}^{t}\left[ K_{B_{1}(t-\tau )}u(\tau ,x)\right] ^{p}d\tau .
\label{iter}
\end{eqnarray}%
From (\ref{weak solution}) and (\ref{eqn-3}), we see that
\begin{eqnarray}
u(t,x) &\geq &K_{t}\varphi (x)+\int_{0}^{t}K_{\tau }f(x)d\tau  \notag \\
&\geq &A\left( K_{B_{1}t}\varphi (x)+tK_{B_{1}t}f(x)\right) .
\label{estimate-1}
\end{eqnarray}%
Starting from (\ref{estimate-1}), we shall apply (\ref{iter}) repeatedly to
deduce the desired inequality (\ref{nonexistence-bound}). Indeed, we obtain
from (\ref{iter}) and (\ref{estimate-1}) that, using the semigroup property (%
\ref{semi:2}) of $\left\{ K_{t}\right\} _{t\geq 0}$ and the elementary
inequality $(a+b)^{p}\geq a^{p}+b^{p}$ for all $p\geq 1$ and $a,b\geq 0$,
\begin{eqnarray*}
u(t,x) &\geq &A\int_{0}^{t}\left[ K_{B_{1}(t-\tau )}u(\tau ,x)\right]
^{p}d\tau \\
&\geq &A\int_{0}^{t}\left[ K_{B_{1}(t-\tau )}\left\{ A\left( K_{B_{1}\tau
}\varphi +\tau K_{B_{1}\tau }f\right) \right\} (x)\right] ^{p}d\tau \\
&=&A^{p+1}\int_{0}^{t}[K_{B_{1}t}\varphi (x)+\tau K_{B_{1}t}f(x)]^{p}d\tau \\
&\geq &A^{p+1}\left\{ t\left( K_{B_{1}t}\varphi (x)\right)
^{p}+\int_{0}^{t}\tau ^{p}(K_{B_{1}t}f(x))^{p}d\tau \right\} \\
&=&A^{p+1}\left\{ t\left( K_{B_{1}t}\varphi (x)\right) ^{p}+\frac{1}{1+p}%
t^{1+p}(K_{B_{1}t}f(x))^{p}\right\} .
\end{eqnarray*}%
Repeating the above procedure, we obtain that for all $n\geq 1,$
\begin{eqnarray*}
u(t,x) &\geq &A^{1+p+\cdots +p^{n}}\left\{ \frac{t^{1+p+\cdots
+p^{n-1}}[K_{Bt}\varphi (x)]^{p^{n}}}{(1+p)^{p^{n-2}}(1+p+p^{2})^{p^{n-3}}%
\cdots (1+p+\cdots +p^{n-1})}\right. \\
&&+\left. \frac{t^{1+p+\cdots +p^{n}}[K_{Bt}f(x)]^{p^{n}}}{%
(1+p)^{p^{n-1}}(1+p+p^{2})^{p^{n-2}}\cdots (1+p+\cdots +p^{n})}\right\} .
\end{eqnarray*}%
It follows that
\begin{align}
A^{(p^{n+1}-1)/(p-1)p^{n}}& t^{(p^{n}-1)/(p-1)p^{n}}K_{Bt}\varphi (x)  \notag
\\
\leq & u(t,x)^{p^{-n}}\prod_{i=2}^{n}(1+p+\cdots +p^{i-1})^{p^{-i}},
\label{bound1} \\
A^{(p^{n+1}-1)/(p-1)p^{n}}& t^{(p^{n+1}-1)/(p-1)p^{n}}K_{Bt}f(x)  \notag \\
\leq & u(t,x)^{p^{-n}}\prod_{i=1}^{n}(1+p+\cdots +p^{i})^{p^{-i}}.
\label{bound2}
\end{align}%
Since
\begin{eqnarray*}
\log \prod_{i=2}^{n}(1+p+\cdots +p^{i-1})^{p^{-i}} &\leq &\sum_{i=2}^{\infty
}\frac{1}{p^{i}}\log (ip^{i})<+\infty , \\
\log \prod_{i=1}^{n}(1+p+\cdots +p^{i})^{p^{-i}} &\leq &\sum_{i=1}^{\infty }%
\frac{1}{p^{i}}\log ((i+1)p^{i})<+\infty ,
\end{eqnarray*}%
and that $u(t,x)$ is essentially bounded on $(0,T)\times M$, we pass to the
limit as $n\rightarrow \infty $ in (\ref{bound1}) and (\ref{bound2}), and
conclude that
\begin{eqnarray}
t^{1/(p-1)}K_{Bt}\varphi (x) &\leq &C_{1}/2,  \label{varphi-bound} \\
t^{p/(p-1)}K_{Bt}f(x) &\leq &C_{1}/2,  \label{f-bound}
\end{eqnarray}%
for some $C_{1}>0.$ Adding (\ref{varphi-bound}) and (\ref{f-bound}), we
obtain (\ref{nonexistence-bound}).
\end{proof}

We are now in a position to obtain the main results of this section.

\begin{theorem}
\label{Non-existence1}Assume that the heat kernel $k$ satisfies conditions $%
(k6)$ and (\ref{general1}). Then the problem (\ref{eq})-(\ref{eq:init}) does
not have any essentially bounded global solution in each of the following
cases:

\begin{enumerate}
\item[(i)] {if $p<1+\frac{\beta }{\alpha }$ and if either $\varphi
(x)\gvertneqq 0$ or $f(x)\gvertneqq 0$.}

\item[(ii)] {if $\alpha \leq \beta $ and if $f(x)\gvertneqq 0$;}

\item[(iii)] {if $\alpha >\beta $ and $p<\frac{\alpha }{\alpha -\beta }(>1+%
\frac{\beta }{\alpha })$ and if $f(x)\gvertneqq 0$. }
\end{enumerate}
\end{theorem}

\begin{proof}
We prove the results by contradiction. Assume that $u(t,x)$ is a
non-negative essentially bounded global solution. Replacing $B_{1}t$ by $t$,
we see from (\ref{nonexistence-bound}) that for all $x\in M$ and $t>0,$
\begin{equation}
t^{1/(p-1)}K_{t}\varphi (x)+t^{p/(p-1)}K_{t}f(x)\leq C_{1},
\label{simple-form}
\end{equation}%
where $0<C_{1}< \infty$ is independent of $\varphi $ and $f$.

\textit{Proof of Case (i):} If $\varphi (x)\gvertneqq 0$, we see from $(k6)$%
, using Fatou's lemma, that
\begin{eqnarray*}
\liminf_{t\rightarrow \infty }\,t^{\alpha /\beta }K_{t}\varphi (x) &\geq
&\liminf_{t\rightarrow \infty }\int_{M}\Phi _{1}\left( \frac{d(x,y)}{%
t^{1/\beta }}\right) \varphi (y)d\mu (y) \\
&\geq &C_{2}.
\end{eqnarray*}%
where $C_{2}=1$ if $||\varphi ||_{1}=\infty $, and $C_{2}=\Phi
_{1}(0)||\varphi ||_{1}$ if $||\varphi ||_{1}<\infty $. However, as $%
1/(p-1)>\alpha /\beta $, this is impossible by using (\ref{simple-form}).
Hence, (\ref{eq})-(\ref{eq:init}) does not have any global essentially
bounded solution.

If $f(x)\gvertneqq 0$, observe that $u(t+t_{0},x)$ is a weak solution of (%
\ref{weak solution}) with initial data $\varphi (x)=u(t_{0},x)$. We may find
$t_{0}>0$ such that $u(t_{0},x)\gvertneqq 0$. Repeating the above argument,
we again see that (\ref{eq})-(\ref{eq:init}) does not have any global
essentially bounded solution.

\textit{Proof of Case (ii):} Observe that by (\ref{weak solution}) and (\ref%
{harnack2}),%
\begin{equation}
u(t,x)\geq \int_{0}^{t}K_{\tau }f(x)d\tau \geq A_{2}tK_{B_{1}t}f(x).
\label{eqn-4}
\end{equation}

We distinguish two cases: $\alpha <\beta $ and $\alpha =\beta .$

$\bullet $ The case $\alpha <\beta $ . It follows from (\ref{eqn-4}) and $%
(k6)$ that%
\begin{eqnarray}
\liminf_{t\rightarrow \infty }t^{(\alpha/\beta)-1}u(t,x) &\geq
&A_{2}\liminf_{t\rightarrow \infty }t^{\alpha/\beta}K_{B_{1}t}f(x)  \notag \\
&\geq &A_{2}\liminf_{t\rightarrow \infty }t^{\alpha/\beta}\int_{M}\frac{1}{%
(B_{1}t)^{\alpha /\beta }}\Phi _{1}\left( \frac{d(x,y)}{(B_{1}t)^{1/\beta }}%
\right) f(y)d\mu (y)  \notag \\
&\geq &C_{3},  \label{case2-1}
\end{eqnarray}%
where $C_{3}=1$ if $||f||_{1}=\infty $ and $C_{3}=A_{2}B_{1}^{-\alpha /\beta
}\Phi _{1}(0)>0$ if $||f||_{1}<\infty $. However, since $u$ is globally
essentially bounded and $\alpha/\beta<1,$ we see%
\begin{equation*}
\liminf_{t\rightarrow \infty }t^{(\alpha/\beta)-1}u(t,x)=0,
\end{equation*}%
a contradiction.

$\bullet $ The case $\alpha =\beta $. For $t>1,$ it follows from ($k6$), ( %
\ref{general1}) and the monotonicity of $\Phi _{2}$ that%
\begin{eqnarray}
u(t,x) &\geq &\int_{0}^{t}K_{\tau }f(x)d\tau  \notag \\
&\geq &\int_{0}^{t}d\tau \int_{M}\tau ^{-1}\Phi _{1}\left( \frac{d(x,y)}{%
\tau ^{1/\beta }}\right) f(y)d\mu (y)  \notag \\
&\geq &a_{1}\int_{1}^{t}d\tau \int_{M}\tau ^{-1}\Phi _{2}\left( a_{2}\frac{%
d(x,y)}{\tau ^{1/\beta }}\right) f(y)d\mu (y)  \notag \\
&\geq &a_{1}\int_{1}^{t}\tau ^{-1}d\tau \int_{M}\Phi
_{2}(a_{2}d(x,y))f(y)d\mu (y).  \label{case2-2}
\end{eqnarray}%
Since {$f(x)\gvertneqq 0$}, we can find a point $x\in M$ such that
\begin{equation*}
\int_{M}\Phi _{2}(a_{2}d(x,y))f(y)d\mu (y)>0.
\end{equation*}%
Passing to the limit as $t\rightarrow \infty $ in (\ref{case2-2}) this
contradicts that $u$ is globally essentially bounded.

\textit{Proof of Case (iii):} It follows from (\ref{simple-form}) and ($k6$)
that
\begin{eqnarray*}
\liminf_{t\rightarrow \infty }C_{1}t^{\alpha /\beta \,-\,p/(p-1)} &\geq
&\liminf_{t\rightarrow \infty }t^{\alpha /\beta }K_{t}f(x) \\
&\geq &\liminf_{t\rightarrow \infty }\int_{M}\Phi _{1}\left( \frac{d(x,y)}{%
t^{1/\beta }}\right) f(y)d\mu (y) \\
&\geq &C_{4},
\end{eqnarray*}%
where $C_{4}=1$ if $||f||_{1}=\infty $ and $C_{4}=\Phi _{1}(0)||f||_{1}$ if $%
||f||_{1}<\infty $. However, this is impossible since $\frac{\alpha }{\beta }%
-\frac{p}{p-1}<0$. The proof is complete.
\end{proof}

In Theorem \ref{Non-existence1}, we do not know in general if there exists any
essentially bounded global solution for two the critical cases $p=1+\frac{%
\beta }{\alpha }$ $(\alpha ,\beta >0)$ and $p=\frac{\alpha }{\alpha -\beta }$
$(\alpha >\beta >0)$.

However,
Theorem \ref{Non-existence1} $(i)$ may be improved to include the critical
exponent $p=1+\frac{\beta }{\alpha }$ under further assumptions {(\ref%
{general4}) and }(\ref{general2}) on the heat kernel $k.$ We first need the
following property.

\begin{proposition}
If $\Phi _{2}$ satisfies (\ref{general4}), then for all $t>0$ and all $%
x,y\in M$,
\begin{equation}
\frac{\Phi _{2}\left( d(x,y)t^{-1/\beta }\right) }{\Phi _{2}\left(
b_{2}d(x,0)t^{-1/\beta }\right) }\geq b_{1}\Phi _{2}\left(
b_{3}d(y,0)t^{-1/\beta }\right) ,  \label{general3}
\end{equation}%
where the constants $b_{i}\, (i=1,2,3)$ are as in (\ref{general4}).
\end{proposition}

\begin{proof}
Since $\Phi _{2}$ is strictly positive and decreasing on $[0,\infty )$ and $%
d(x,y)\leq d(x,0)+d(y,0),$ we have
\begin{equation}
\Phi _{2}\left( d(x,y)t^{-1/\beta }\right) \geq \Phi _{2}\left(
d(x,0)t^{-1/\beta }+d(y,0)t^{-1/\beta }\right) .  \label{monotone}
\end{equation}%
It follows from (\ref{general4}) that
\begin{equation*}
\Phi _{2}\left( d(x,0)t^{-1/\beta }+d(y,0)t^{-1/\beta }\right) \geq
b_{1}\Phi _{2}\left( b_{2}d(x,0)t^{-1/\beta }\right) \Phi _{2}\left(
b_{3}d(y,0)t^{-1/\beta }\right) ,
\end{equation*}%
which combines with (\ref{monotone}) to give (\ref{general3}).
\end{proof}

\begin{theorem}
\label{T-nonexis2} Assume that the heat kernel $k$ satisfies conditions $%
(k5),(k6)$ and (\ref{general1}), {(\ref{general4}) and }(\ref{general2}).
Then (\ref{eq})-(\ref{eq:init}) does not have any essentially bounded global
solutions if $p\leq 1+\frac{\beta }{\alpha }$ and if either $\varphi
(x)\gvertneqq 0$ or $f(x)\gvertneqq 0$.
\end{theorem}

\begin{proof}
In view of Theorem \ref{Non-existence1} $(i)$ it is enough to consider the
critical exponent $p=1+\beta /\alpha .$ We only consider the case $\varphi
(x)\gvertneqq 0$ (the case $f(x)\gvertneqq 0$ may be treated in a similar
way). Then (\ref{simple-form}) becomes
\begin{equation*}
t^{\alpha /\beta }K_{t}\varphi (x)+t^{1+\alpha /\beta }K_{t}f(x)\leq C_{1}.
\end{equation*}%
From condition $(k6)$
\begin{equation}
\int_{M}\varphi (y)d\mu (y)\leq C_{2},  \label{case1-0}
\end{equation}%
where $C_{2}=C_{1}/\Phi _{1}(0)$. For any $t_{0}>0$, the function $%
v(t,x)\equiv u(t+t_{0},x)$ is a weak solution to (\ref{weak solution}) with
initial data $\varphi (x)=u(t_{0},x)$. Repeating the procedure of (\ref%
{case1-0}), we have that for all $t_{0}>0,$
\begin{equation}
\int_{M}u(t_{0},y)d\mu (y)\leq C_{2}.  \label{case1-1}
\end{equation}

We claim that there exist positive constants $\gamma ,\rho $ possibly
depending on $t_{0}$ and $\varphi $ such that, for all $x\in M,$
\begin{equation}
u(t_{0},x)\geq \rho k(\gamma ,x,0).  \label{case1-2}
\end{equation}%
To see this, observe that
\begin{equation*}
\Phi _{2}\left( d(x,0)\gamma ^{-1/\beta }\right) \geq k(\gamma ,x,0)\gamma
^{\alpha /\beta },
\end{equation*}%
and thus, using (\ref{general3}) and setting $\gamma =(a_{1}b_{2})^{-\beta
}t_{0},$
\begin{eqnarray*}
\Phi _{2}\left( a_{2}d(x,y)t_{0}^{-1/\beta }\right) &\geq &b_{1}\Phi
_{2}\left( a_{1}b_{3}d(y,0)t_{0}^{-1/\beta }\right) \Phi _{2}\left(
a_{1}b_{2}d(x,0)t_{0}^{-1/\beta }\right) \\
&\geq &b_{1}\Phi _{2}\left( a_{1}b_{3}d(y,0)t_{0}^{-1/\beta }\right)
k(\gamma ,x,0)\gamma ^{\alpha /\beta }.
\end{eqnarray*}%
Using (\ref{weak solution}) and (\ref{general1}),
\begin{eqnarray*}
u(t_{0},x) &\geq &\int_{M}k(t_{0},x,y)\varphi (y)d\mu (y) \\
&\geq &t_{0}^{-\alpha /\beta}\int_{M}\Phi _{1}\left( d(x,y)t_{0}^{-1/\beta
}\right) \varphi (y)d\mu (y) \\
&\geq &a_{1}t_{0}^{-\alpha /\beta}\int_{M}\Phi _{2}\left(
a_{2}d(x,y)t_{0}^{-1/\beta }\right) \varphi (y)d\mu (y) \\
&\geq &a_{1}b_{1}\left( \frac{\gamma }{t_{0}}\right) ^{\alpha
/\beta}k(\gamma ,x,0)\int_{M}\Phi _{2}\left( a_{1}b_{3}d(y,0)t_{0}^{-1/\beta
}\right) \varphi (y)d\mu (y),
\end{eqnarray*}%
and hence, inequality (\ref{case1-2}) holds by setting%
\begin{equation*}
\rho :=a_{1}b_{1}\left( \frac{\gamma }{t_{0}}\right) ^{\alpha /\beta
}\int_{M}\Phi _{2}\left( a_{1}b_{3}d(y,0)t_{0}^{-1/\beta }\right) \varphi
(y)d\mu (y),
\end{equation*}%
proving our claim.

Consider $v(t,x)\equiv u(t+t_{0},x)$ such that $u(t_{0},x)\gvertneqq 0$.
Applying (\ref{case1-2}), we obtain
\begin{eqnarray*}
v(t,x) &\geq &\int_{M}k(t,x,y)u(t_{0},y)d\mu (y) \\
&\geq &\rho \int_{M}k(t,x,y)k(\gamma ,y,0)d\mu (y) \\
&=&\rho k(t+\gamma ,x,0),
\end{eqnarray*}%
which yields that, using (\ref{weak solution}), $(k5)$ and Fubini's theorem,
\begin{eqnarray}
\int_{M}v(t,x)d\mu (x) &\geq &\int_{M}d\mu (x)\int_{0}^{t}d\tau
\int_{M}k(t-\tau ,x,y)v^{p}(\tau ,y)d\mu (y)  \notag \\
&=&\int_{0}^{t}d\tau \int_{M}v^{p}(\tau ,y)d\mu (y)  \notag \\
&\geq &\rho ^{p}\int_{0}^{t}d\tau \int_{M}k^{p}(\tau +\gamma ,y,0)d\mu (y).
\label{case1-3}
\end{eqnarray}%
As $p=1+\beta /\alpha $, we see from (\ref{general2}) and $(k6)$ that
\begin{eqnarray*}
k^{p}(\tau +\gamma ,y,0) &\geq &(\tau +\gamma )^{-(1+\alpha /\beta )}\Phi
_{1}^{p}\left( d(y,0)(\tau +\gamma )^{-1/\beta }\right) \\
&\geq &c_{1}(\tau +\gamma )^{-(1+\alpha /\beta )}\Phi _{2}\left(
c_{2}d(y,0)(\tau +\gamma )^{-1/\beta }\right) \\
&=&c_{1}c_{2}^{-\alpha }(\tau +\gamma )^{-1}[c_{2}^{-\beta }(\tau +\gamma
)]^{-\alpha /\beta }\Phi _{2}\left( c_{2}d(y,0)(\tau +\gamma )^{-1/\beta
}\right) \\
&\geq &c_{1}c_{2}^{-\alpha }(\tau +\gamma )^{-1}k(c_{2}^{-\beta }(\tau
+\gamma ),y,0),
\end{eqnarray*}%
which combines with (\ref{case1-3}) to give that
\begin{equation}
\int_{M}v(t,x)d\mu (x)\geq c_{1}c_{2}^{-\alpha }\rho ^{p}\int_{0}^{t}(\tau
+\gamma )^{-1}d\tau .  \label{case1-4}
\end{equation}%
Passing to the limit as $t\rightarrow \infty $, we conclude that
\begin{equation*}
\int_{M}v(t,x)d\mu (x)\rightarrow \infty ,
\end{equation*}%
which contradicts (\ref{case1-1}).
\end{proof}

We note that our results agree with the earlier ones where $M=\mathbb{R}^{n}$
and $\mu $ is Lebesgue measure, and where the heat kernel $k$ is the
Gauss-Weierstrass function (so that $\Delta $ is the usual Laplacian), see
\cite{Fujita66, Wei80, Wei81, Bandle00}. See also \cite{FalHu01} where $M$
is a fractal and $\mu $ is $\alpha $-dimenisonal Hausdorff measure, and
where $k$ is the Gauss-type heat kernel on $M.$

\section{Existence of solutions}

\label{exis} In this section we give sufficient conditions for local
existence and global existence of weak solutions.

\begin{theorem}[Local-existence]
\label{existence} Suppose that the heat kernel $k$ satisfies $(k6)$. Let $%
b(t)$ be a continuously differentiable function on $[0,T_{0})$ satisfying
\begin{equation}
b^{\prime }(t)=b^{p}(t)\left[ \int_{0}^{t}\frac{||K_{\tau }f||_{\infty }}{%
b(\tau )}d\tau +||K_{t}\varphi ||_{\infty }\right] ^{p-1}  \label{b(t)-ode}
\end{equation}%
with initial value $b(0)=1$. If
\begin{equation}
\int_{0}^{T_{0}}\left[ \int_{0}^{s}\frac{||K_{\tau }f||_{\infty }}{b(\tau )}%
d\tau +||K_{s}\varphi ||_{\infty }\right] ^{p-1}ds\leq \frac{1}{p-1},
\label{existence-estimate}
\end{equation}%
then (\ref{eq})-(\ref{eq:init}) has a non-negative local solution $u\in
L^{\infty }((0,T),M)$ for all $0< T<T_{0}$, provided that $
||\varphi ||_{\infty }<\infty $.
\end{theorem}

\begin{remark}
\label{R-exis}By Peano's theorem, there exists some $T_{0}>0$ and some
continuous differentiable function $b(t)$ such that (\ref{b(t)-ode}) holds
in $[0,T_{0})$. Clearly, such a $b(t)$ is non-decreasing in $[0,T_{0})$. On
the other hand, condition (\ref{existence-estimate}) may be verified for
some specific cases. For example, if $\left( k5\right) $ holds and if $%
f=0,\varphi =C>0,$ then
\begin{equation*}
b(t)=[1-(p-1)C^{p-1}t]^{-1/(p-1)}
\end{equation*}%
satisfies (\ref{b(t)-ode}) in $[0,T_{0})$ where $T_{0}=(p-1)^{-1}C^{-(p-1)}$%
, and (\ref{existence-estimate}) also holds. As an another example, let $%
f=1,\varphi =0$ and assume $\left( k5\right) $ holds. Then, for $p=2,$ we
see that $b(t)=1/\cos t$ satisfies (\ref{b(t)-ode}) for $t\in \lbrack 0,\pi
/2)$, and that (\ref{existence-estimate}) holds.
\end{remark}

\begin{proof}
Define
\begin{equation*}
a(t)=b(t)\int_{0}^{t}\frac{||K_{\tau }f||_{\infty }}{b(\tau )}d\tau .
\end{equation*}
Note that $a(0)=0$ and $a(t)\geq 0$ for $t\in \lbrack 0,T_{0})$. 
Incorporating this into  (\ref{b(t)-ode}), we get 
\begin{equation*}
b^{\prime }(t) =b(t)\left[ a(t)+b(t)||K_{t}\varphi ||_{\infty }\right]
^{p-1}  \label{b(t)}.
\end{equation*}
Moreover, 
\begin{eqnarray*}
a^{\prime }(t) &=&  ||K_{t}f||_{\infty } + \frac{b'(t)a(t)}{b(t)}\nonumber\\
&=& ||K_{t}f||_{\infty }+a(t)\left[ a(t)+b(t)||K_{t}\varphi
||_{\infty }\right] ^{p-1}.
\end{eqnarray*}
Together with the initial conditions, these differential equations are equivalent to 
\begin{eqnarray}
a(t) &=&\int_{0}^{t}||K_{\tau }f||_{\infty }d\tau +\int_{0}^{t}a(\tau
)(a(\tau )+b(\tau )||K_{\tau }\varphi ||_{\infty })^{p-1}d\tau ,
\label{a(t)-b(t)1} \\
b(t) &=&1+\int_{0}^{t}b(\tau )(a(\tau )+b(\tau )||K_{\tau }\varphi
||_{\infty })^{p-1}d\tau .  \label{a(t)-b(t)2}
\end{eqnarray}

Let $
\mathcal{H}$ be the family of continuous functions $u$ satisfying
\begin{equation}
K_{t}\varphi (x)\leq u(t,x)\leq a(t)+b(t)K_{t}\varphi (x)\ \text{ for all }%
(t,x)\in \lbrack 0,T_{0})\times M.  \label{sub-sup}
\end{equation}
Define
\begin{equation}
\mathcal{F}u(t,x)=K_{t}\varphi (x)+\int_{0}^{t}K_{\tau }f(x)d\tau
+\int_{0}^{t}K_{t-\tau }u^{p}(\tau ,x)d\tau .  \label{def-1}
\end{equation}%
We claim that if $u\in \mathcal{H}$, then $\mathcal{F}u\in
\mathcal{H}$, that is,
\begin{equation}
K_{t}\varphi (x)\leq \mathcal{F}u(t,x)\leq a(t)+b(t)K_{t}\varphi (x)\quad
(0\leq t <T_{0}, x \in  M).  \label{def-2}
\end{equation}%
Observe that, using $(k1)$,%
\begin{eqnarray*}
&&\hspace{-1.5cm} \int_{0}^{t}K_{t-\tau }[a(\tau )+b(\tau )K_{\tau }\varphi ]^{p}(x)d\tau \\
&=&\int_{0}^{t}d\tau \int_{M}k(t-\tau ,x,y)\left[ a(\tau )+b(\tau )K_{\tau
}\varphi (y)\right] ^{p}d\mu (y) \\
&\leq &\int_{0}^{t}[a(\tau )+b(\tau )||K_{\tau }\varphi ||_{\infty }]^{p-1}
\left[ a(\tau )+b(\tau )K_{t}\varphi (x)\right] d\tau .
\end{eqnarray*}%
It follows from (\ref{def-1}) and (\ref{sub-sup}) that
\begin{eqnarray*}
\mathcal{F}u(t,x) &\leq &K_{t}\varphi (x)+\int_{0}^{t}K_{\tau }f(x)d\tau
+\int_{0}^{t}K_{t-\tau }[a(\tau )+b(\tau )K_{\tau }\varphi ]^{p}(x)d\tau
\notag \\
&\leq &\left[ \int_{0}^{t}||K_{\tau }f||_{\infty }d\tau +\int_{0}^{t}a(\tau
)[a(\tau )+b(\tau )||K_{\tau }\varphi ||_{\infty }]^{p-1}d\tau \right]
\notag \\
&&+\left[ 1+\int_{0}^{t}b(\tau )[a(\tau )+b(\tau )||K_{\tau }\varphi
||_{\infty }]^{p-1}d\tau \right] K_{t}\varphi (x)\\
&=& a(t)+b(t)K_{t}\varphi (x)
\end{eqnarray*}
using (\ref{a(t)-b(t)1}) and  (\ref{a(t)-b(t)2}),
so (\ref{def-2}) holds,
proving our claim.

For $n=0,1,2,\cdots ,$ define
\begin{eqnarray*}
u_{0}(t,x) &=&K_{t}\varphi (x), \\
u_{n+1}(t,x) &=&\mathcal{F}u_{n}(t,x).
\end{eqnarray*}%
Using (\ref{def-1}) inductively, it follows that the sequence $%
\{u_{n}(t,x)\} $ is non-decreasing in $n$, and, for all $n\geq 0$ and all $%
x\in M,t\in \lbrack 0,T_{0}),$ satisfies
\begin{equation*}
K_{t}\varphi (x)\leq u_{n}(t,x)\leq a(t)+b(t)K_{t}\varphi (x).
\end{equation*}%
Let $u(t,x):=\lim\limits_{n\rightarrow \infty }u_{n}(t,x)$. Note that $%
K_{t}\varphi (x)\leq u(t,x)\leq a(t)+b(t)K_{t}\varphi (x)$. Using the
monotone convergence theorem, we have
\begin{equation*}
\lim_{n\rightarrow \infty }\int_{0}^{t}d\tau \int_{M}k(t-\tau
,x,y)u_{n}^{p}(\tau ,y)d\mu (y)=\int_{0}^{t}d\tau \int_{M}k(t-\tau
,x,y)u^{p}(\tau ,y)d\mu (y).
\end{equation*}%
Since $u_{n}(t,x)$ satisfies
\begin{equation}
u_{n+1}(t,x)=K_{t}\varphi (x)+\int_{0}^{t}K_{\tau }f(x)d\tau
+\int_{0}^{t}K_{t-\tau }u_{n}^{p}(\tau ,x)d\tau ,  \label{limits}
\end{equation}%
we pass to the limit as $n\rightarrow \infty $ to obtain
\begin{equation*}
u(t,x)=K_{t}\varphi (x)+\int_{0}^{t}K_{\tau }f(x)d\tau
+\int_{0}^{t}K_{t-\tau }u^{p}(\tau ,x)d\tau ,
\end{equation*}%
which shows that $u(t,x)$ is a non-negative local solution of (\ref{eq})-(%
\ref{eq:init}) for $t\in \lbrack 0,T_{0})$.

Since $a(t)$, $b(t)$ are differentiable functions on $[0,T_{0})$, we see
from (\ref{sub-sup}) that for all $t\in \lbrack 0,T_{0})$,
\begin{equation*}
||u(t,\cdot )||_{\infty }\leq ||a(t)+b(t)K_{t}\varphi ||_{\infty }<\infty .
\end{equation*}%
The proof is complete.
\end{proof}

Recall that, by Theorem \ref{Non-existence1}, (\ref{eq})-(\ref{eq:init}) does
not have any essentially bounded global weak solution {if $\alpha >\beta $
and $p<\frac{\alpha }{\alpha -\beta }$ and if $f(x)\gvertneqq 0$}. However,
we can show that (\ref{eq})-(\ref{eq:init}) possesses a essentially bounded
global solution if {$p>\frac{\alpha }{\alpha -\beta },$ for small functions }%
$f$ and $\varphi $ (cf. \cite{zhang98} for Euclidean spaces). 
To do
this, we need some integral estimates which are consequences of measure bounds for small and large balls.


Recall that a measure  $\mu $ on a metric measure space  is {\it upper $\alpha $-regular} if there exist some $C,\alpha >0$
such that
\begin{equation}
\mu (B(x,r))\leq Cr^{\alpha }\text{ \ (for all }x\in M,r>0\text{),}  \label{upper reg}
\end{equation}
and is $\alpha $-\emph{regular} if there exists a
constant $C>0$ such that for all $x\in M$ and all $r>0,$
\begin{equation}
C^{-1}r^{\alpha }\leq \mu (B(x,r))\leq Cr^{\alpha }\text{ \ (for all }x\in M,r>0\text{).}
  \label{volume}
\end{equation}%
It was shown in \cite[Theorem 3.2]{GHL03} that if the heat
kernel $k$ satisfies $(k5),(k6)$ with $\Phi _{2}(s)$ satisfying%
\begin{equation}
\int_{0}^{\infty }s^{\alpha -1}\Phi _{2}(s)ds<\infty ,  \label{phi}
\end{equation}%
then the measure $\mu $ is $\alpha $-\emph{regular}.
Note that, by the monotonicity of $\Phi _{2},$ condition (\ref{phi}) implies
that $s^{\alpha }\Phi _{2}(s)\leq C<\infty $ for all $s\in \lbrack 0,\infty
).$

\begin{proposition}
\label{sup-radial} Assume that $\mu $ is upper $\alpha $-regular and $x_{0}$
is a reference point in $M$. If $0 < \lambda _{1}< \alpha $ and $\lambda
_{1}+\lambda _{2}>\alpha ,$ then there exists a constant $C_{0}>0$ such that%
\begin{equation}
\int_{M}\frac{1}{d(y,x)^{\lambda _{1}}[1+d(y,x_{0})^{\lambda
_{2}}]}d\mu (y)\leq C_{0}\quad \text{{\rm (for all }}x\in M).  \label{sup-dis}
\end{equation}
\end{proposition}

\begin{proof} 
For each $x\in M,$ let $\Omega _{1}=\left\{ y\in M:d(y,x)\geq
d(y,x_{0})\right\} $ and $\Omega _{2}=M\setminus \Omega _{1}.$ Then
\begin{equation*}
\int_{\Omega _{1}}\frac{1}{d(y,x)^{\lambda _{1}}[1+d(y,x_{0})^{\lambda _{2}}]
}d\mu (y) 
\leq \int_{M}\frac{1}{d(y,x_{0})^{\lambda _{1}}[1+d(y,x_{0})^{\lambda
_{2}}]}d\mu (y)
\end{equation*}
and
\begin{equation*}
\int_{\Omega _{2}}\frac{1}{d(y,x)^{\lambda _{1}}[1+d(y,x_{0})^{\lambda _{2}}]
}d\mu (y) \leq \int_{M}\frac{1}{d(y,x)^{\lambda _{1}}[1+d(y,x)^{\lambda _{2}}]}d\mu
(y).
\end{equation*}
Routine estimates using upper regularity (\ref{upper reg}) now give uniform bounds on these integrals near $x_0$ and $x$ (since $ \lambda _{1}< \alpha $) and for large $d(y,x_{0})$ and $d(y,x)$ (since $\lambda
_{1}+\lambda _{2}>\alpha$), to give (\ref{sup-dis}).
\end{proof}

\begin{proposition}
Assume that $\mu $ is upper $\alpha $-regular and $x_{0}$ is a reference
point in $M$. If $0< \lambda _{1}<\alpha $ and $\lambda _{2}>\alpha ,$
then there exists a constant $C_{1}>0$ such that%
\begin{equation}
\int_{M}\frac{1}{d(y,x)^{\lambda _{1}}[1+d(y,x_{0})^{\lambda _{2}}]}d\mu
(y)\leq \frac{C_{1}}{1+d(x,x_{0})^{\lambda _{1}}}.  \label{eqn-10}
\end{equation}
\end{proposition}

\begin{proof}
\bigskip Fix $x\in M.$ If $d(x,x_{0})\leq 1,$ then (\ref{eqn-10}) directly
follows from (\ref{sup-dis}), since%
\begin{equation*}
\int_{M}\frac{1}{d(y,x)^{\lambda _{1}}[1+d(y,x_{0})^{\lambda _{2}}]}d\mu
(y)\leq C_{0}\leq \frac{2C_{0}}{1+d(x,x_{0})^{\lambda _{1}}}.
\end{equation*}

Assume that $d(x,x_{0})\geq 1$. If $d(y,x_{0})\geq d(x,x_{0})/2$, we have
that
\[
\frac{1}{1+d(y,x_{0})^{\lambda _{2}}}\leq \frac{C}{[1+d(y,x_{0})^{\lambda
_{2}-\lambda _{1}}]\left[ 1+d(x,x_{0})^{\lambda _{1}}\right] }
\]%
where $C$ is independent of $x_{0},y$. Using Proposition \ref{sup-radial},
it follows that
\begin{eqnarray}
&&\int_{d(y,x_{0})\geq d(x,x_{0})/2}\frac{1}{d(y,x)^{\lambda
_{1}}[1+d(y,x_{0})^{\lambda _{2}}]}d\mu (y)  \nonumber \\
&\leq &\frac{C}{1+d(x,x_{0})^{\lambda _{1}}}\int_{d(y,x_{0})\geq
d(x,x_{0})/2}\frac{1}{d(y,x)^{\lambda _{1}}[1+d(y,x_{0})^{\lambda
_{2}-\lambda _{1}}]}d\mu (y)  \nonumber \\
&\leq &\frac{C_{1}}{1+d(x,x_{0})^{\lambda _{1}}}.  \label{eqn-11}
\end{eqnarray}%
If $d(y,x_{0})<d(x,x_{0})/2,$ then
\begin{equation}
d(y,x)^{-\lambda _{1}} \leq \left[ d(x,x_{0})-d(y,x_{0})\right] ^{-\lambda
_{1}} 
\leq \left[ d(x,x_{0})/2\right] ^{-\lambda _{1}}\leq \frac{2^{\lambda
_{1}+1}}{1+d(x,x_{0})^{\lambda _{1}}},
\end{equation}
and hence,
\begin{equation}
\int_{d(y,x_{0})<d(x,x_{0})/2}\frac{1}{d(y,x)^{\lambda
_{1}}[1+d(y,x_{0})^{\lambda _{2}}]}d\mu (y)\leq \frac{C_{2}}{%
1+d(x,x_{0})^{\lambda _{1}}}.  \label{eqn-12}
\end{equation}%
where we have used that $\int_{M}\frac{1}{1+d(y,x_{0})^{\lambda _{2}}}%
d\mu (y)<\infty $ as $\lambda _{2}>\alpha $.

Adding (\ref{eqn-11}) and (\ref{eqn-12}) we see that
 (\ref{eqn-10}) also holds if $d(x,x_{0})\geq 1.$
\end{proof}

We now show the global existence of weak solutions for small $\varphi $ and $%
f.$

\begin{theorem}[Global-existence]
\label{global} Let $\alpha >\beta >0$ and suppose that the heat kernel $k$ satisfies $(k5),(k6)$ 
and that $\Phi _{2}$ satisfies (\ref{phi}). Let $\lambda >\alpha $ and let $x_{0}$ be a reference point in $M$.
The for each $p>\alpha/(\alpha -\beta)$ there exists  $\delta >0$ such that if
\begin{equation*}
0<\varphi (x),f(x)\leq \frac{\delta }{1+d(x,x_{0})^{\lambda }}
\end{equation*}
then (\ref{eq})-(\ref{eq:init}) has an essentially bounded
global solution.
\end{theorem}

\begin{proof}
Recall that conditions $(k5),(k6)$ and (\ref{phi}) imply that $\mu $ is $
\alpha $-regular. Let the map $\mathcal{F}$ be defined as in (\ref{def-1}):
\begin{equation*}
\mathcal{F}u(t,x)=K_{t}\varphi (x)+\int_{0}^{t}K_{\tau }f(x)d\tau
+\int_{0}^{t}K_{t-\tau }u^{p}(\tau ,x)d\tau .
\end{equation*}%
For $\epsilon >0$, let $S_{\varepsilon }$ be the complete subset of the Banach space 
$L^{\infty }\left( \lbrack 0,\infty
)\times M \right)$ given by
\begin{equation*}
S_{\varepsilon }=\left\{ u\in L^{\infty }\left( \lbrack 0,\infty
)\times M \right) :0\leq u(t,x)\leq \frac{\varepsilon }{1+d(x,x_{0})^{\alpha -\beta }}
\right\}
\end{equation*}
 We
will use the contraction principle to show that, for appropriately small $\epsilon$ and $\delta$,  there exists a global
solution in $S_{\varepsilon }$.

For $\lambda>\alpha$, we
claim that there exists $C_{2}>0$ such that, for all  $0\leq g(x)\leq \delta/(1+d(x,x_{0})^{\lambda })$, we have
\begin{equation}
K_{t}g(x)\leq \frac{C_{2}\delta }{1+d(x,x_{0})^{^{\alpha }}}\text{ for all }
x\in M\text{ and all }t>0.  \label{Kg}
\end{equation}
To see this, let $x\in M.$ If $d(x,x_{0})\leq 1$, then (\ref{Kg}) is
clear since
\begin{eqnarray*}
K_{t}g(x) &=&\int_{M}k(t,x,y)g(y)d\mu (y) \\
&\leq &\int_{M}\frac{\delta }{1+d(y,x_{0})^{\lambda }}k(t,x,y)d\mu (y) \\
&\leq &\delta \int_{M}k(t,x,y)d\mu (y)\leq \delta \\
&\leq &\frac{2\delta }{1+d(x,x_{0})^{^{\alpha }}}.
\end{eqnarray*}%
So assume  $d(x,x_{0})>1.$ We have, using condition $\left( k6\right)
, $
\begin{eqnarray}
K_{t}g(x) &\leq &\int_{M}\frac{\delta }{1+d(y,x_{0})^{\lambda }}k(t,x,y)d\mu
(y)  \notag \\
&\leq &\delta \left\{ \int_{\Omega _{1}}\frac{1}{1+d(y,x_{0})^{\lambda }}
\frac{1}{t^{\alpha /\beta }}\Phi _{2}\left( \frac{d(y,x)}{t^{1/\beta }}
\right) d\mu (y)\right.  \notag \\
&&\quad +\left. \int_{\Omega _{2}}\frac{1}{1+d(y,x_{0})^{\lambda }}k(t,x,y)d\mu
(y)\right\},  \label{eqn-30}
\end{eqnarray}
where $\Omega _{1}=\left\{ y\in M:d(y,x_{0})\leq d(x,x_{0})/2\right\} $ and $
\Omega _{2}=M\setminus \Omega _{1}.$ For $y\in \Omega _{1},$ we have, noting from (\ref{phi}) that $s^{\alpha} \Phi_2(s)$ is bounded, 
\begin{eqnarray*}
\frac{1}{t^{\alpha /\beta }}\Phi _{2}\left( \frac{d(y,x)}{t^{1/\beta }}
\right) &=&\frac{1}{d(y,x)^{\alpha }}\left( \frac{d(y,x)}{t^{1/\beta }}
\right) ^{\alpha }\Phi _{2}\left( \frac{d(y,x)}{t^{1/\beta }}\right) \\
&\leq &\frac{C}{d(y,x)^{\alpha }}\leq \frac{2^{\alpha }C}{d(x,x_{0})^{\alpha
}} \\
&\leq &\frac{2^{\alpha +1}C}{1+d(x,x_{0})^{\alpha }},
\end{eqnarray*}%
and hence, using that $\int_{M}\frac{d\mu (y)}{1+d(y,x_{0})^{
\lambda }}<+\infty $ for $\lambda >\alpha ,$
\begin{eqnarray}
\int_{\Omega _{1}}\frac{1}{1+d(y,x_{0})^{\lambda }}\frac{1}{t^{\alpha /\beta
}}\Phi _{2}\left( \frac{d(y,x)}{t^{1/\beta }}\right) d\mu (y) &\leq &\frac{%
2^{\alpha +1}C}{1+d(x,x_{0})^{\alpha }}\int_{\Omega _{1}}\frac{d\mu (y)}{%
1+d(y,x_{0})^{\lambda }}  \notag \\
&\leq &\frac{C}{1+d(x,x_{0})^{\alpha }}.  \label{eqn-31}
\end{eqnarray}
For $y\in \Omega _{2}$,
\begin{eqnarray}
\int_{\Omega _{2}}\frac{1 }{1+d(y,x_{0})^{\lambda }}k(t,x,y)d\mu (y) &\leq &%
\frac{2^{\lambda }}{1+d(x,x_{0})^{\lambda }}\int_{\Omega _{2}}k(t,x,y)d\mu
(y)  \notag \\
&\leq &\frac{C}{1+d(x,x_{0})^{\alpha }}.  \label{eqn-32}
\end{eqnarray}%
using  that $\lambda >\alpha$. Adding (\ref{eqn-31}) and (\ref{eqn-32}), we see that (\ref{Kg}) follows
from (\ref{eqn-30}), proving our claim.

Observe that by $\left( k6\right) $ and (\ref{phi}),%
\begin{eqnarray}
\int_{0}^{t}k(\tau ,x,y)d\tau &\leq &\int_{0}^{t}\frac{1}{\tau ^{\alpha
/\beta }}\Phi _{2}\left( \frac{d(y,x)}{\tau ^{1/\beta }}\right) d\tau  \notag
\\
&=&\frac{\beta }{d(y,x)^{\alpha -\beta }}\int_{d(x,y)/t^{1/\beta }}^{\infty
}s^{\alpha -\beta -1}\Phi _{2}(s)ds  \notag \\
&\leq &\frac{\beta }{d(y,x)^{\alpha -\beta }}\int_{0}^{\infty }s^{\alpha
-\beta -1}\Phi _{2}(s)ds  \notag \\
&\leq &\frac{C}{d(y,x)^{\alpha -\beta }},  \label{eqn-33}
\end{eqnarray}%
since
\begin{equation*}
\int_{0}^{\infty }s^{\alpha -\beta -1}\Phi _{2}(s)ds
\leq \Phi _{2}(0)\int_{0}^{1}s^{\alpha -\beta -1}ds+\int_{1}^{\infty
}s^{\alpha -1}\Phi _{2}(s)ds 
<+\infty,
\end{equation*}%
using the monotonicity of $\Phi_2$ and (\ref{phi}).

Therefore,  using (\ref{eqn-33}) and (\ref{eqn-10}) with $
\lambda _{1}=\alpha -\beta >0$ and $\lambda _{2}=\lambda >\alpha ,$
\begin{eqnarray}
\int_{0}^{t}K_{\tau }f(x)d\tau &=&\int_{M}\left[ \int_{0}^{t}k(\tau,x,y)d%
\tau \right] f(y)d\mu (y)  \notag \\
&\leq &\int_{M}\frac{C}{d(y,x)^{\alpha -\beta }}\frac{\delta }{
\big(1+d(y,x_{0})^{\lambda }\big)}d\mu (y)  \notag \\
&\leq &\frac{C\delta }{1+d(x,x_{0})^{\alpha -\beta }}  \label{eqn-35}
\end{eqnarray}%
for all $x\in M$ and $t>0.$ Similarly, for $u\in S_{\varepsilon },$ we have
that, using (\ref{eqn-10}) with $\lambda _{1}=\alpha -\beta ,\lambda
_{2}=p(\alpha -\beta )>\alpha ,$
\begin{eqnarray}
\int_{0}^{t}K_{t-\tau }u^{p}(\tau ,x)d\tau &\leq
&\int_{0}^{t}\int_{M}k(t-\tau ,x,y)\frac{\varepsilon ^{p}}{%
(1+d(y,x_{0})^{\alpha -\beta })^{p}}d\mu (y)d\tau  \notag \\
&\leq &\int_{M}\frac{C}{d(y,x)^{\alpha -\beta }}\frac{\varepsilon ^{p}}{%
(1+d(y,x_{0})^{\alpha -\beta })^{p}}d\mu (y)  \notag \\
&\leq &C\varepsilon ^{p}\int_{M}\frac{1}{d(y,x)^{\alpha -\beta }}\frac{1}{%
1+d(y,x_{0})^{(\alpha -\beta )p}}d\mu (y)  \notag \\
&\leq &\frac{C\varepsilon ^{p}}{1+d(x,x_{0})^{\alpha -\beta }}
\label{eqn-36}
\end{eqnarray}%
for all $x\in M$ and $t>0.$ It follows from (\ref{Kg}), (\ref%
{eqn-35}), (\ref{eqn-36}) that if $u\in S_{\varepsilon },$ then
\begin{eqnarray*}
\mathcal{F}u(t,x) &\leq &\frac{C_{2}\delta }{1+d(x,x_{0})^{^{\alpha }}}+%
\frac{C\delta+C\varepsilon ^{p}}{1+d(x,x_{0})^{\alpha -\beta }} \\
&\leq &\frac{C_{1}(\delta +\varepsilon ^{p})}{1+d(x,x_{0})^{\alpha -\beta }}
\\
&\leq &\frac{\varepsilon }{1+d(x,x_{0})^{\alpha -\beta }}
\end{eqnarray*}%
provided that $C_{1}\left( \delta +\varepsilon ^{p}\right) \leq \varepsilon $,
in which case $\mathcal{F}S_{\varepsilon }\subset S_{\varepsilon }$.

Next we show that $\mathcal{F}$ is contractive on $S_{\varepsilon }.$
Indeed, for $u_{1},u_{2}\in S_{\varepsilon }$, we have
\begin{equation*}
\left\vert \mathcal{F}u_{1}(t,x)-\mathcal{F}u_{2}(t,x)\right\vert \leq
\int_{0}^{t}\int_{M}k(t-\tau ,x,y)\left\vert u_{1}^{p}(\tau
,y)-u_{2}^{p}(\tau ,y)\right\vert d\mu (y)d\tau .
\end{equation*}%
Using the elementary inequality
\begin{equation*}
|a^{p}-b^{p}|\leq p\max \{a^{p-1},b^{p-1}\}|a-b|\text{ for }a,b\geq 0,p>1,
\end{equation*}%
and the definition of $S_{\varepsilon }$, we obtain, using (\ref%
{eqn-33}) and (\ref{sup-dis}), that
\begin{eqnarray*}
&&\hspace{-1.5cm}\left\vert \mathcal{F}u_{1}(t,x)-\mathcal{F}u_{2}(t,x)\right\vert  \\
&\leq &||u_{1}-u_{2}||_{\infty}\int_{0}^{t}\int_{M}k(t-\tau ,x,y)\frac{p\varepsilon
^{p-1}}{[1+d(y,x_{0})^{\alpha -\beta }]^{p-1}}d\mu (y)d\tau  \\
&\leq &||u_{1}-u_{2}||_{\infty}\int_{M}\frac{C}{d(y,x)^{\alpha -\beta }}\frac{%
p\varepsilon ^{p-1}}{1+d(y,x_{0})^{\left( \alpha -\beta \right) (p-1)}}d\mu
(y) \\
&\leq &C_{3}p\varepsilon ^{p-1}||u_{1}-u_{2}||_{\infty}.
\end{eqnarray*}%
Thus if $\epsilon$ is  small enough to ensure that both $C_{3}p\varepsilon ^{p-1}<1$ and $C_{1} \varepsilon ^{p} < \varepsilon $, and then $\delta$ is chosen small enough so that 
$C_{1}\left( \delta +\varepsilon ^{p}\right) \leq \varepsilon $, 
applying Banach's contraction principle to $\mathcal{F}$ on
the complete set $S_{\varepsilon }$
implies that (\ref{weak solution}) and thus (\ref{eq})-(\ref{eq:init})
has a global positive solution in $S_{\varepsilon }$.
\end{proof}

\section{Regularity}

\label{regular} In this section, we discuss the regularity of weak
solutions. We show that weak solutions are H\"{o}lder continuous in the
spatial variable $x$ if the source term $f$ and initial value $\varphi $ are
both H\"{o}lder continuous\textbf{. }We adapt the method used in \cite%
{FalHu01}.

In order to obtain the regularity of weak solutions, we need to assume that
the function $\Phi _{2}$ in condition $(k6)$ satisfies the following
assumption:
\begin{equation}
\int_{0}^{\infty }s^{\alpha }\Phi _{2}(s)ds<\infty ,\text{ }
\label{general5}
\end{equation}%
where $\alpha $ is as in condition $(k6).$ Since $\Phi _{2}$ is
non-increasing on $[0,\infty ),$ condition (\ref{general5}) implies that $
s^{1+\alpha }\Phi _{2}(s)=o(1)$ as $s\rightarrow \infty .$

Clearly, the Gauss-type function $\Phi _{2}$ defined as in (\ref{G-func})
satisfies condition (\ref{general5}) for all $\gamma >0$ whilst the
Cauchy-type function $\Phi _{2}$ defined as in (\ref{C-func}) satisfies
condition (\ref{general5}) for all $\gamma >1+\alpha .$

Note that condition (\ref{general5}) is stronger than (\ref{phi}%
), and hence\ it implies that $\mu $ is $\alpha $-regular$.$

\begin{proposition}
\label{int-estimate} Assume that $\mu $ is upper $\alpha $-regular. If $\Phi
_{2}$ satisfies (\ref{general5}) then, for all $\lambda \in (0,1],$
\begin{equation}
\int_{M}d(x,y)^{\lambda }\Phi _{2}\left( \frac{d(x,y)}{t^{1/\beta }}\right)
d\mu (y)\leq C_{2}t^{(\alpha +\lambda )/\beta } \quad \text{ \ (for all }x\in M,t>0\text{)}
 \label{int}
\end{equation}%
for some constant $C_{2}$.
\end{proposition}

\begin{proof}
Let $g(r)=r^{\lambda }\Phi _{2}\left( \frac{r}{t^{1/\beta }}\right) $ for $
r>0.$ From (\ref{int}) $g(r) = o(r^{-\alpha})$ so, by a standard argument  
using $\alpha$-regularity and integration by parts (see \cite[Proposition 4.1]{FalHu01}), it follows  that
\begin{align*}
\int_{M}d(x,y)^{\lambda }& \Phi _{2}\left( \frac{d(x,y)}{t^{1/\beta }}
\right) d\mu (y)
=  \int_{M} g\big(d(x,y)\big) d\mu (y)\\
& \leq C_{1}\int_{0}^{\infty }r^{\alpha }|g^{\prime }(r)|dr \\
& =C_{1}\int_{0}^{\infty }r^{\alpha }\left\vert \lambda r^{\lambda -1}\Phi
_{2}\left( \frac{r}{t^{1/\beta }}\right) +r^{\lambda }\Phi _{2}^{\prime
}\left( \frac{r}{t^{1/\beta }}\right) t^{-1/\beta }\right\vert dr \\
& \leq C_{2}t^{(\alpha +\lambda )/\beta }\left[ \int_{0}^{\infty }\lambda
s^{\alpha +\lambda -1}\Phi _{2}(s)ds+\int_{0}^{\infty }s^{\alpha +\lambda
}\left( -\Phi _{2}^{\prime }(s)\right) ds\right] .
\end{align*}
By an easy calculation, the last integral
\begin{eqnarray*}
\int_{0}^{\infty }s^{\alpha +\lambda }\left( -\Phi _{2}^{\prime }(s)\right)
ds &=&-\left. s^{\alpha +\lambda }\Phi _{2}(s)\right\vert _{0}^{\infty
}+(\alpha +\lambda )\int_{0}^{\infty }s^{\alpha +\lambda -1}\Phi _{2}(s)ds \\
&=&(\alpha +\lambda )\int_{0}^{\infty }s^{\alpha +\lambda -1}\Phi
_{2}(s)ds\leq C_{3}
\end{eqnarray*}%
using (\ref{general5}). Therefore,
\begin{equation*}
\int_{M}d(x,y)^{\lambda }\Phi \left( \frac{d(x,y)}{t^{1/\beta }}\right) d\mu
(y)\leq C_{2}t^{(\alpha +\lambda )/\beta },
\end{equation*}%
as desired.
\end{proof}

We now show the H\"{o}lder continuity of weak solutions of (\ref{weak
solution}).

\begin{theorem}[H\"{o}lder Continuity]
Assume that $\varphi $, $f\in L^{1}(M)$ are H\"{o}lder continuous with
exponents $\theta _{1},\theta _{2}\in (0,1]$ respectively: for all $%
x_{1},x_{2}\in M,$
\begin{eqnarray}
\left\vert \varphi (x_{1})-\varphi (x_{2})\right\vert &\leq
&C_{5}d(x_{1},x_{2})^{\theta _{1}},  \label{varphi-holder} \\
\left\vert f(x_{1})-f(x_{2})\right\vert &\leq &C_{6}d(x_{1},x_{2})^{\theta
_{2}},  \label{f-holder}
\end{eqnarray}%
where $C_{5},C_{6}>0$. Assume that the heat kernel $k$ satisfies $(k5)-(k7)$
and that $\Phi _{2}$ satisfies (\ref{general5}) with $\lambda =\max \left\{
\theta _{1},\theta _{2}\right\} $. Let $u(t,x)$ be a non-negative weak
solution to (\ref{eq})-(\ref{eq:init}) that is bounded in $(0,T)\times M$
for some $T>0$. Then $u(t,x)$ is H\"{o}lder continuous: for all $%
x_{1},x_{2}\in M$ and all $t\in (0,T),$
\begin{equation}
|u(t,x_{1})-u(t,x_{2})|\leq Cd(x_{1},x_{2})^{\theta },  \label{u-cts}
\end{equation}%
where $\theta =\theta _{1}\sigma /(\theta _{1}+\nu \beta )$ and $C>0$ may
depend on $T$ but is independent of $t,x$.
\end{theorem}

\begin{proof}
From $(k6)$, (\ref{varphi-holder}), and (\ref{int}), there exists $C>0$ such
that for all $t>0$ and $x\in M,$
\begin{eqnarray}
\int_{M}k(t,x,y)\left\vert \varphi (y)-\varphi (x)\right\vert d\mu (y) &\leq
&C_{5}t^{-\alpha /\beta }\int_{M}d(x,y)^{\theta _{1}}\Phi _{2}\left( \frac{%
d(x,y)}{t^{1/\beta }}\right) d\mu (y)  \notag \\
&\leq &Ct^{\theta _{1}/\beta }.  \label{eqn-7}
\end{eqnarray}%
By (\ref{weak solution}) it is enough to show that each of the functions $%
u_{0},u_{1},u_{2}$ is H\"{o}lder continuous in $(0,T)\times M$, where
\begin{eqnarray*}
u_{0}(t,x) &=&K_{t}\varphi (x), \\
u_{1}(t,x) &=&\int_{0}^{t}K_{\tau }f(x)d\tau , \\
u_{2}(t,x) &=&\int_{0}^{t}K_{t-\tau }u^{p}(\tau ,x)d\tau .
\end{eqnarray*}%
We first show the H\"{o}lder continuity of $u_{0}$. Indeed, for $t>0$ and $%
x_{1},x_{2}\in M,$ we see from $(k7)$ that
\begin{eqnarray}
|u_{0}(t,x_{1})-u_{0}(t,x_{2})| &=&\left\vert
\int_{M}(k(t,x_{1},y)-k(t,x_{2},y))\varphi (y)d\mu (y)\right\vert  \notag \\
&\leq &Lt^{-\nu }d(x_{1},x_{2})^{\sigma }||\varphi ||_{1}  \notag \\
&\leq &L||\varphi ||_{1}d(x_{1},x_{2})^{\sigma -\nu s_{0}}
\label{u0-estimate1}
\end{eqnarray}%
if $t\geq d(x_{1},x_{2})^{s_{0}}$, where $s_{0}>0$ will be specified later
on. On the other hand, if $t\leq d(x_{1},x_{2})^{s_{0}}$, we have, using $%
(k5)$, (\ref{eqn-7}) and (\ref{varphi-holder}), that
\begin{eqnarray*}
|u_{0}(t,x_{1})-u_{0}(t,x_{2})| &\leq &\left\vert
\int_{M}k(t,x_{1},y)(\varphi (y)-\varphi (x_{1}))d\mu (y)\right. \\
&&\left. +\left[ \varphi (x_{1})-\varphi (x_{2})\right] -%
\int_{M}k(t,x_{2},y)(\varphi (y)-\varphi (x_{2}))d\mu (y)\right\vert \\
&\leq &2Ct^{\theta _{1}/\beta }+C_{5}d(x_{1},x_{2})^{\theta _{1}} \\
&\leq &C\left[ d(x_{1},x_{2})^{s_{0}\theta _{1}/\beta
}+d(x_{1},x_{2})^{\theta _{1}}\right] .
\end{eqnarray*}%
Combining this with (\ref{u0-estimate1}), it follows that
\begin{eqnarray}
|u_{0}(t,x_{1})-u_{0}(t,x_{2})| &\leq &C\left[ d(x_{1},x_{2})^{\sigma -\nu
s_{0}}+d(x_{1},x_{2})^{s_{0}\theta _{1}/\beta }+d(x_{1},x_{2})^{\theta _{1}}%
\right]  \notag \\
&\leq &Cd(x_{1},x_{2})^{\theta _{1}\sigma /(\theta _{1}+\nu \beta )},
\label{u0-holder1}
\end{eqnarray}%
for all $t>0$ and $x_{1},x_{2}\in M$ with $d(x_{1},x_{2})\leq 1$, where $%
s_{0}=\sigma /\left( \nu +\frac{\theta _{1}}{\beta }\right) $ so that $%
\sigma -\nu s_{0}=s_{0}\theta _{1}/\beta $, and where we have used the fact
that $\theta _{1}\geq s_{0}\theta _{1}/\beta $ for $\sigma \leq 1\leq \nu $
and $\beta \geq 1$.

Next we show the H\"{o}lder continuity of $u_{1}$. As with (\ref{eqn-7}), we
have from $(k6)$, (\ref{f-holder}) and (\ref{int}) that%
\begin{equation*}
\int_{M}k(\tau ,x_{1},y)\left\vert f(y)-f(x_{1})\right\vert d\mu (y)\leq
C\tau ^{\theta _{2}/\beta },
\end{equation*}%
which yields that, using $(k5)$ and (\ref{f-holder}),
\begin{eqnarray}
\left\vert u_{1}(t,x_{1})-u_{1}(t,x_{2})\right\vert &=&\left\vert
\int_{0}^{t}\left[ K_{\tau }f(x_{1})-K_{\tau }f(x_{2})\right] d\tau
\right\vert  \notag \\
&=&\left\vert \int_{0}^{t}d\tau \int_{M}k(\tau ,x_{1},y)(f(y)-f(x_{1}))d\mu
(y)\right.  \notag \\
&&\left. +t\left[ f(x_{1})-f(x_{2})\right] -\int_{0}^{t}d\tau \int_{M}k(\tau
,x_{2},y)(f(y)-f(x_{2}))d\mu (y)\right\vert  \notag \\
&\leq &2C\int_{0}^{t}\tau ^{\theta _{2}/\beta }d\tau
+C_{6}td(x_{1},x_{2})^{\theta _{2}}  \notag \\
&=&Ct^{\theta _{2}/\beta +1}+C_{6}td(x_{1},x_{2})^{\theta _{2}}  \notag \\
&\leq &C\left[ d(x_{1},x_{2})^{s_{1}+s_{1}\theta _{2}/\beta
}+d(x_{1},x_{2})^{s_{1}+\theta _{2}}\right]  \label{u1-estimate1}
\end{eqnarray}%
if $t\leq d(x_{1},x_{2})^{s_{1}}$, where $s_{1}>0$ will be chosen later.

On the other hand, if $t>d(x_{1},x_{2})^{s_{1}},$ and setting $%
t_{1}=d(x_{1},x_{2})^{s_{1}}$, we obtain, using $(k7)$, that
\begin{eqnarray}
\left\vert \int_{t_{1}}^{t}\left[ K_{\tau }f(x_{1})-K_{\tau }f(x_{2})\right]
d\tau \right\vert &\leq &\int_{t_{1}}^{t}d\tau \int_{M}\left\vert k(\tau
,x_{1},y)-k(\tau ,x_{2},y)\right\vert \left\vert f(y)\right\vert d\mu (y)
\notag \\
&\leq &\int_{t_{1}}^{t}L\tau ^{-\nu }d(x_{1},x_{2})^{\sigma }||f||_{1}d\tau
\notag \\
&\leq &L\frac{t_{1}^{1-\nu }-t^{1-\nu }}{\nu -1}d(x_{1},x_{2})^{\sigma
}||f||_{1}  \notag \\
&\leq &\frac{L}{\nu -1}d(x_{1},x_{2})^{s_{1}(1-\nu )+\sigma }||f||_{1}.
\label{Pf-estimate1}
\end{eqnarray}%
It follows from (\ref{Pf-estimate1}) and (\ref{u1-estimate1}) that
\begin{eqnarray}
\left\vert u_{1}(t,x_{1})-u_{1}(t,x_{2})\right\vert &\leq &\left\vert
\int_{0}^{t_{1}}K_{\tau }f(x_{1})-K_{\tau }f(x_{2})d\tau \right\vert
+\left\vert \int_{t_{1}}^{t}K_{\tau }f(x_{1})-K_{\tau }f(x_{2})d\tau
\right\vert  \notag \\
&\leq &C\left[ d(x_{1},x_{2})^{s_{1}+s_{1}\theta _{2}/\beta
}+d(x_{1},x_{2})^{s_{1}+\theta _{2}}+d(x_{1},x_{2})^{s_{1}(1-\nu )+\sigma }%
\right]  \notag \\
&\leq &Cd(x_{1},x_{2})^{\sigma (\theta _{2}+\beta )/(\theta _{2}+\nu \beta )}
\label{u1-holder1}
\end{eqnarray}%
if $d(x_{1},x_{2})\leq 1$, where $s_{1}=\sigma \beta /(\theta _{2}+\nu \beta
)$ so that $s_{1}+s_{1}\theta _{2}/\beta =s_{1}(1-\nu )+\sigma $, and where
we have used the fact that
\begin{equation*}
s_{1}+\theta _{2}\geq s_{1}+s_{1}\theta _{2}/\beta
\end{equation*}%
for $\sigma \leq 1\leq \nu $ and $\beta \geq 1$.

Finally, we show the H\"{o}lder continuity of $u_{2}.$ Since $u(t,x)$ is
bounded on $(0,T)\times M$, we see that
\begin{equation*}
\int_{t-\eta }^{t}d\tau \int_{M}k(t-\tau ,x,y)u^{p}(\tau ,y)d\mu (y)\leq
C\eta .
\end{equation*}%
Hence, using $(k7)$, we obtain
\begin{eqnarray*}
\left\vert u_{2}(t,x_{1})-u_{2}(t,x_{2})\right\vert &=&\left\vert
\int_{t-\eta }^{t}d\tau \int_{M}k(t-\tau ,x_{1},y)u^{p}(\tau ,y)d\mu
(y)\right. \\
&&-\int_{t-\eta }^{t}d\tau \int_{M}k(t-\tau ,x_{2},y)u^{p}(\tau ,y)d\mu (y)
\\
&&+\left. \int_{0}^{t-\eta }d\tau \int_{M}(k(t-\tau ,x_{1},y)-k(t-\tau
,x_{2},y))u^{p}(\tau ,y)d\mu (y)\right\vert \\
&\leq &2C\eta +L\int_{0}^{t-\eta }d\tau \int_{M}|t-\tau |^{-\nu
}d(x_{1},x_{2})^{\sigma }u^{p}(\tau ,y)d\mu (y) \\
&\leq &C(\eta +\eta ^{1-\nu }d(x_{1},x_{2})^{\sigma }).
\end{eqnarray*}%
Taking $\eta =d(x_{1},x_{2})^{\sigma /\nu }$, we thus have
\begin{equation}
\left\vert u_{2}(t,x_{1})-u_{2}(t,x_{2})\right\vert \leq
Cd(x_{1},x_{2})^{\sigma /\nu }.  \label{u2-holder1}
\end{equation}%
Combining (\ref{u0-holder1}), (\ref{u1-holder1}) and (\ref{u2-holder1}), we
conclude that
\begin{equation*}
|u(t,x_{1})-u(t,x_{2})|\leq Cd(x_{1},x_{2})^{\theta _{1}\sigma /(\theta
_{1}+\nu \beta )}.
\end{equation*}%
for all $t\in (0,T)$ and $x_{1},x_{2}\in M$ with $d(x_{1},x_{2})\leq 1,$ for
some $C>0$, where we have used that
\begin{equation*}
\theta _{1}\sigma /(\theta _{1}+\nu \beta )\leq \sigma /\nu \leq \sigma
(\theta _{2}+\beta )/(\theta _{2}+\nu \beta ).
\end{equation*}%
The proof is complete.
\end{proof}

Finally, one may show that if the heat kernel $k$
satisfies $(k5)$, if $\left\Vert f\right\Vert _{\infty }<\infty $ and if $%
\varphi (x)$ satisfies%
\begin{equation*}
|K_{t+\delta }\varphi (x)-K_{t}\varphi (x)|\leq C\delta \quad \text{ (for all } t>0,x\in M),
\end{equation*}%
then the essentially bounded weak solution $u$ of (\ref%
{weak solution}) is Lipschitz continuous\ in time $t$ on $(0,T)\times M,$
that is,%
\begin{equation*}
|u(t+\delta ,x)-u(t,x)|\leq C_{1}\delta \quad (t\in (0,T), \delta >0, x\in M).
\end{equation*}%
 We omit the details, which are similar to the special case considered in \cite{FalHu01}.

We note that, unlike the blow-up and the existence, the
regularity of solutions is not related to the Hausdorff dimension $\alpha $
and the walk dimension $\beta .$

\end{document}